# Simultaneous mapping of magnetic and atomic structure for direct visualization of nanoscale magnetoelastic coupling

Sangjun Kang[1, 2], Maximilian Töllner[1], Di Wang[1, 3], Christian Minnert[4], Karsten Durst[4], Arnaud Caron[5], Rafal E. Dunin-Borkowski[6], Jeffrey McCord[7], Christian Kübel[1, 2, 3*] and Xiaoke Mu[1, 8*]

[1]*Institute of Nanotechnology (INT), Karlsruhe Institute of Technology (KIT), 76344 Eggenstein-Leopoldshafen, Germany*
[2]*Joint Research Laboratory Nanomaterials, Technical University of Darmstadt (TUDa), 64287 Darmstadt, Germany*
[3]*Karlsruhe Nano Micro Facility (KNMFi), Karlsruhe Institute of Technology (KIT), 76344 Eggenstein-Leopoldshafen, Germany*
[4]*Physical Metallurgy, Department of Materials Science, Technical University of Darmstadt (TUDa), 64287 Darmstadt, Germany*
[5]*Korea University of Technology and Education (Koreatech), 330708 Cheonan, Republic of Korea*
[6]*Ernst Ruska-Centre for Microscopy and Spectroscopy with Electrons, Peter Grünberg Institute, Forschungszentrum Jülich GmbH, 52425 Jülich, Germany.*
[7]*Nanoscale Magnetic Materials – Magnetic Domains, Department of Materials Science, Kiel University, 24143, Kiel, Germany*
[8]*School of Materials and Energy and Electron Microscopy Centre, Lanzhou University, Lanzhou 730000, China*

*Correspondings: xiaoke.mu@kit.edu and christian.kuebel@kit.edu*

*Abstract:*

*Achieving a correlative measurement of both magnetic and atomic structures at the nanoscale is imperative to understand the fundamental magnetism of matters and for fostering the development of new magnetic nanomaterials. Conventional microscopy methods fall short in providing the two information simultaneously. Here, we develop a new approach to simultaneously map the magnetic field and atomic structure at the nanoscale using Lorentz 4-dimensional scanning transmission electron microscopy (Ltz-4D-STEM). This method enables precise measurement of the characteristic atomic and magnetic structures across an extensive field of view, a critical aspect for investigating real-world ferromagnetic materials. It offers a comprehensive visualization and statistical evaluation of the different structural information at a pixel-by-pixel correlation. The new method allows to directly visualize the magnetoelastic coupling and the resulting complex magnetization arrangement as well as the competition between magnetoelastic and magnetostatic energy. This approach opens new avenues for in-depth studying the structure-property correlation of nanoscale magnetic materials.*

**Introduction**

Magnetic microstructures including domain size, domain walls, and vortices play a pivotal role in determining the magnetic properties of materials and significantly contribute to the design of magnetic devices. [1-6] The magnetic structure is intrinsically determined by the interatomic distances and coordination. [7-9] Characterization of the atomic structure of a material and correlating it with the magnetic structures at the nanoscale is essential for unraveling fundamental magnetism and advancing the development of new magnets with tailored and improved performance.

For instance, the manipulation of magnetic structures through strain-controlled adjustments via magnetoelastic coupling has been extensively researched and leads to a wide range of applications in magnetostrictive devices, magnetic sensors, memories, and actuators. [10-13] Despite these advances, the nanoscale relationship between magnetism and strain remains poorly understood due to the current limitations in characterization methods, in which a direct and meaningful comparison of the magnetic and atomic structural information cannot be obtained easily at the same time.

With the advances in electron microscopy, conventional scanning/transmission electron microscopy (S/TEM) can now routinely capture atomic structures at sub-Angstrom resolution. However, S/TEM has challenges in magnetic imaging due to the strong magnetic fields produced by the objective lens in proximity to the sample, which alters the magnetization state of the sample. A field-free Lorentz (Ltz) optical setup, which offers magnetic field-free conditions at the sample position by eliminating the use of the objective lens, has been widely adopted in Lorentz TEM, electron holography, and STEM differential phase contrast (DPC) for imaging the magnetic structure within a sample. [11, 14-16] However, operating commercially available microscopes in this mode comes at the expense of spatial resolution and lacks atomic-level information, even with the availability of aberration correction. [17-20] Recently atomic imaging under field-free conditions has been demonstrated in a specially designed microscope with a newly invented objective lens system. [21] Nevertheless, in practice, this direct real-space atomic imaging limits the field of view to a few tens of nanometers, far less than the typical length scale of magnetic domains or domain walls, which can be micrometers or larger, especially in soft magnetic materials. These challenges prevent a correlative characterization of the local magnetic and atomic structure.

Here, we advanced the optics in field-free mode in a standard microscope and developed a technique to simultaneously map magnetic and atomic structures using four-dimensional (4D)-STEM in the new Lorentz optical setup (Ltz-4D-STEM). Unlike techniques relying on high-resolution atomic lattice imaging, the development of 4D-STEM has provided a means to map atomic structures over a large field of view by recording information in reciprocal space by scanning a nanoscale electron probe across the samples. [22-24] In the conventional field-free mode, the diffraction signal contains limited information about the atomic structure as the high-angle scattered electrons are physically blocked by the projection lenses. Our development enables simultaneous recording of both the unscattered electron beam and diffracted beams representing interatomic pairs, which provides at the same time the phase shift of the electron exit wave induced by the local magnetic field and the information on the atomic structure of the sample. We demonstrate the new method by directly visualizing the local magnetic field and strain coupling in a Fe-based amorphous alloy, which has been of strong interest owing to its magnetic softness together with significant magnetostriction, contributing e.g. to the field of ultrasensitive magnetic field sensors. [1, 25, 26] Using the pixel-to-pixel correlation between the spin and strain field maps, versatile analyses of the mutual interaction between the atomic structure and the local magnetization in the material were achieved.

**Method development**

The development of the new Ltz-4D-STEM mode has been conducted using a Themis Z double-corrected TEM (ThermoFisher Scientific) operated at 300 kV. The approach can be used for all kinds of commercial TEMs equipped with a Lorentz lens and operated at all typical high tensions. Figure 1a schematically shows the Ltz-4D-STEM idea for sample investigation. A quasi-parallel electron probe with a small semi-convergence angle (in our case it is 0.26 mrad and limited by the uncorrected spherical aberration of the condenser system) in microprobe STEM mode is focused (to ~ 5 nm diameter in our case) on an electron transparent sample. The electron diffraction patterns are acquired from the nano-volume at each scan position during stepwise scanning of the probe over the area of interest.

The diffraction data offer rich structural information through reciprocal space. For instance, in the case of crystalline materials, the diffraction spots convey details of the crystal symmetry, lattice parameters, and orientation (refer to Figure 1b, top). For amorphous matter characterized by a set of diffuse rings due to the lack of long-range order (Figure 1b, bottom), the diffraction rings provide information on the short/medium-range atomic arrangement e.g., the inter-atomic distance and atomic coordination. This motivated the application of 4D-STEM for conventional nano-beam electron diffraction (Figure 1c). The pole pieces of the objective lens act as probe- and diffraction-forming lenses, which are less than a few millimeters away from the sample. This enables high-resolution direct imaging and access to high-angle scattered electrons (>50 mrad) for performing structural analysis. To achieve field-free conditions, both the upper and lower pole pieces of the objective lens must be turned off. In the conventional field-free STEM mode (used in all commercial microscopes) shown in Figure 1d, the last condenser lens with a spherical aberration of ten meters is used for forming the probe. This limits the probe diameter and thus the spatial resolution of the scanned image. At the same time, the diffraction lens is used to collect the electrons scattered from the sample and create a diffraction pattern at its back focal plane. As the diffraction lens is located far away from the sample (more than 10 cm for uncorrected systems), the liner tube acts as a physical aperture and blocks diffracted beams at high angles, thus hindering the recording of atomic-level structural information (e.g., the recording angle is limited to < 2.2 mrad, corresponding to observable lattice spacings larger than ~ 0.9 nm in the Themis Z). Only the direct beam and surroundings can be used for the imaging of magnetic domains. [27, 28]

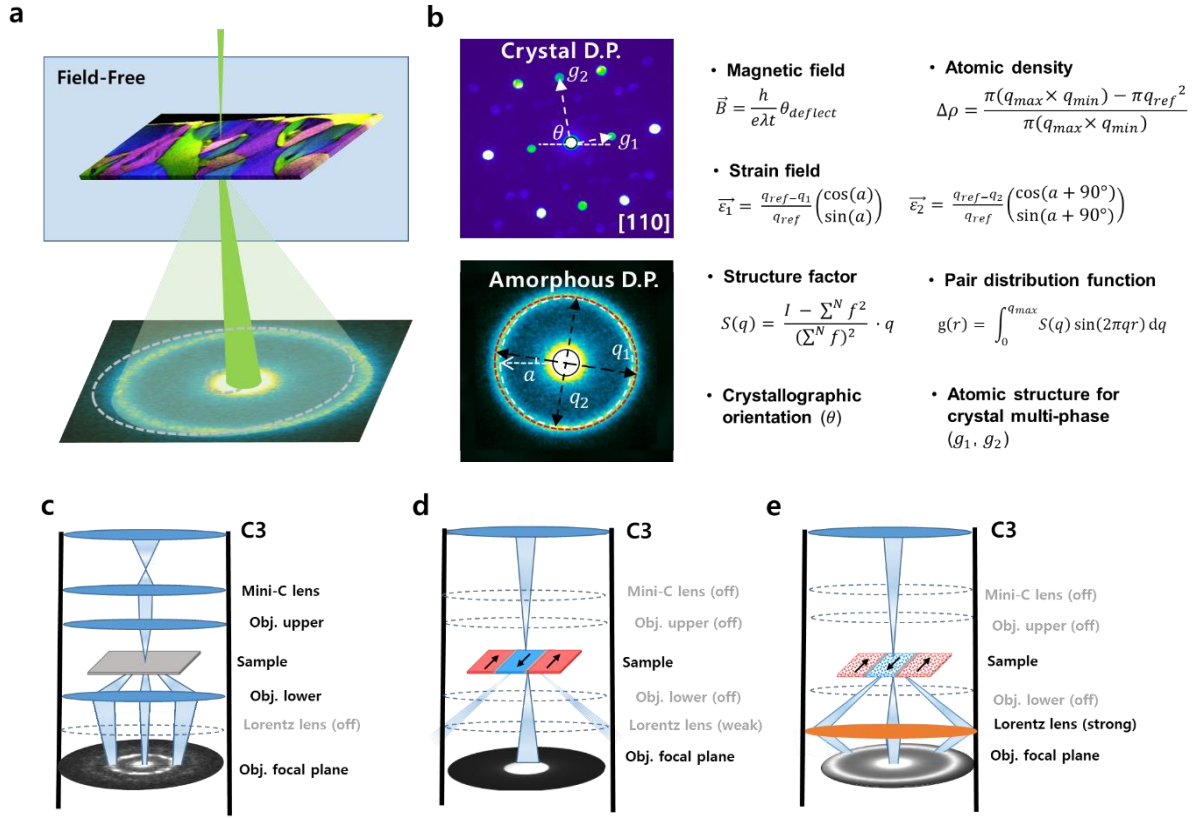

**Fig. 1:** Schematic illustration of Ltz-4D-STEM and its optical setup. **a** The electron probe is focused on the TEM sample in a free-field condition (Lorentz mode). Spatially-resolved diffraction patterns are collected during scanning over an area of interest using 4D-STEM. **b** Exemplary data analysis: The position of the direct beam measures the momentum transferred by the Lorentz force, reflecting the local magnetic field of the sample. The relative atomic density can be calculated by quantifying the area encircled by the 1st diffraction ring. The 1$^{st}$ and 2$^{nd}$ principal strain ($\vec{\varepsilon}_1$ and $\vec{\varepsilon}_2$) can be calculated from the elliptical distortion of the diffraction ring. The diffraction patterns can be processed to determine the structure factor $S(q)$ by azimuthal integration into intensity profiles $I(q)$ and background subtraction. A PDF can be obtained by Fourier sine transformation of $S(q)$ at every scan position. Moreover, Ltz-4D-STEM of crystalline materials can provide information on crystal symmetry, orientation, lattice parameters, etc. **c-e** Illustrations of the optical setup using in conventional nanobeam 4D-STEM, conventional field-free 4D-STEM, and the new Ltz-4D-STEM.

In order to overcome this issue and to get access to highly scattered electrons containing lattice information < 1 Å, we employ the Lorentz lens, designed for regular field-free TEM, in the field-free STEM mode. As depicted in Figure 1e, the Lorentz lens is positioned in close proximity (~ 3 cm) to the sample. This positioning allows the lens to capture highly scattered electrons before they are obstructed by the liner tube, all while ensuring that the magnetic field acting on the sample remains unaffected. In an image aberration-corrected microscope, the

Lorentz lens is integrated with the corrector system as the 1$^{st}$ transfer lens. The two hexapole lenses of the image corrector were switched off to eliminate any 3 and 6-fold distortion, and the diffraction and projection lenses were adjusted to image the post-sample diffraction plane to a camera located in the projection chamber with minimal distortions and optimized magnification of the diffraction pattern (i.e. the camera length) to fit the size of the camera. As a result, the highest accessible diffraction angle in the new setup is increased to about 25 mrad without introducing significant distortions (Figure S1). Thereby, the attainable maximum structural information was enhanced from approximately about 9 Å (in the conventional field-free mode) to < 0.8 Å (in the new mode). Arrays of diffraction patterns containing both unscattered and highly scattered beams can be captured using 4D-STEM with the new lens setup, which we refer to as Ltz-4D-STEM. Figure 1b shows two typical diffraction patterns of crystalline and amorphous materials recorded in this mode, as well as exemplary data analysis. Analysis of all diffraction patterns with both low- and high-angle signals enables simultaneous imaging of magnetic and atomic structures.

The in-plane component of the magnetic fields inside of the sample deflects the electron beam through the Lorentz force (Figure S2a-c). Therefore, the center position of the diffraction pattern at each probe position reflects the direction and strength of the local in-plane magnetic field that the electron probe has passed through. The in-plane magnetic field can be calculated as $\vec{B} = \frac{h}{e\lambda t}\theta_{deflect}$, where $\theta_{deflect}$ is the beam deflection angle, $h$ Planck's constant, $\lambda$ the relativistic electron wavelength, $e$ the charge of an electron, and $t$ the sample thickness, which can be estimated from a thickness map (e.g., Figure S3) obtained by electron energy loss spectroscopy (EELS). For an amorphous diffraction pattern, the center of the diffraction ring can be used to measure the beam deflection providing higher accuracy than using the transmitted center beam directly. In parallel, stress-induced distortions of the atomic structure can be determined for amorphous materials from the azimuthal elliptic distortion of the diffraction rings (Figure S2d-g). The residual elastic strain field within the sample can be mapped by quantifying the elliptical deviation of all diffraction patterns in the 4D-STEM data and comparing it with a reference area, employing the data analysis process established by Gammer et al. [29] and Kang et al. [23]. In this work, we use the convention that the first principal strain is oriented along the compression direction providing the compressive component ($\vec{\varepsilon}_{com}$), which in turn results in the tensile component ($\vec{\varepsilon}_{ten}$) as the second principal strain that is oriented perpendicular to the first principal strain. We also quantify the deviatoric strain from the strain components to disentangle the true local distortion of the material and the local net volume change as a response to the local hydrostatic stress. The diffraction intensity in the radial direction of the nanobeam diffraction pattern reflects the distances between atoms in the nanoscale volume, giving rise to information on the atomic density under the approximation that no considerable chemical variations are present. Following previous works [30, 31], we estimated the relative atomic density by calculating the area encircled by the 1$^{st}$ ring of each diffraction pattern in the 4D-STEM data. This approach takes the elliptical deviation of the diffraction rings due to deviatoric strain into account and is an intuitive way to analyze the volumetric strain and disentangle the density information from the sample thickness. One can further obtain the local pair distribution function (PDF) for providing information on atomic configurations at each scanning position using the same dataset. In this work, we used the relative atomic density to describe the atomic packing for ease of presentation.

**Results**

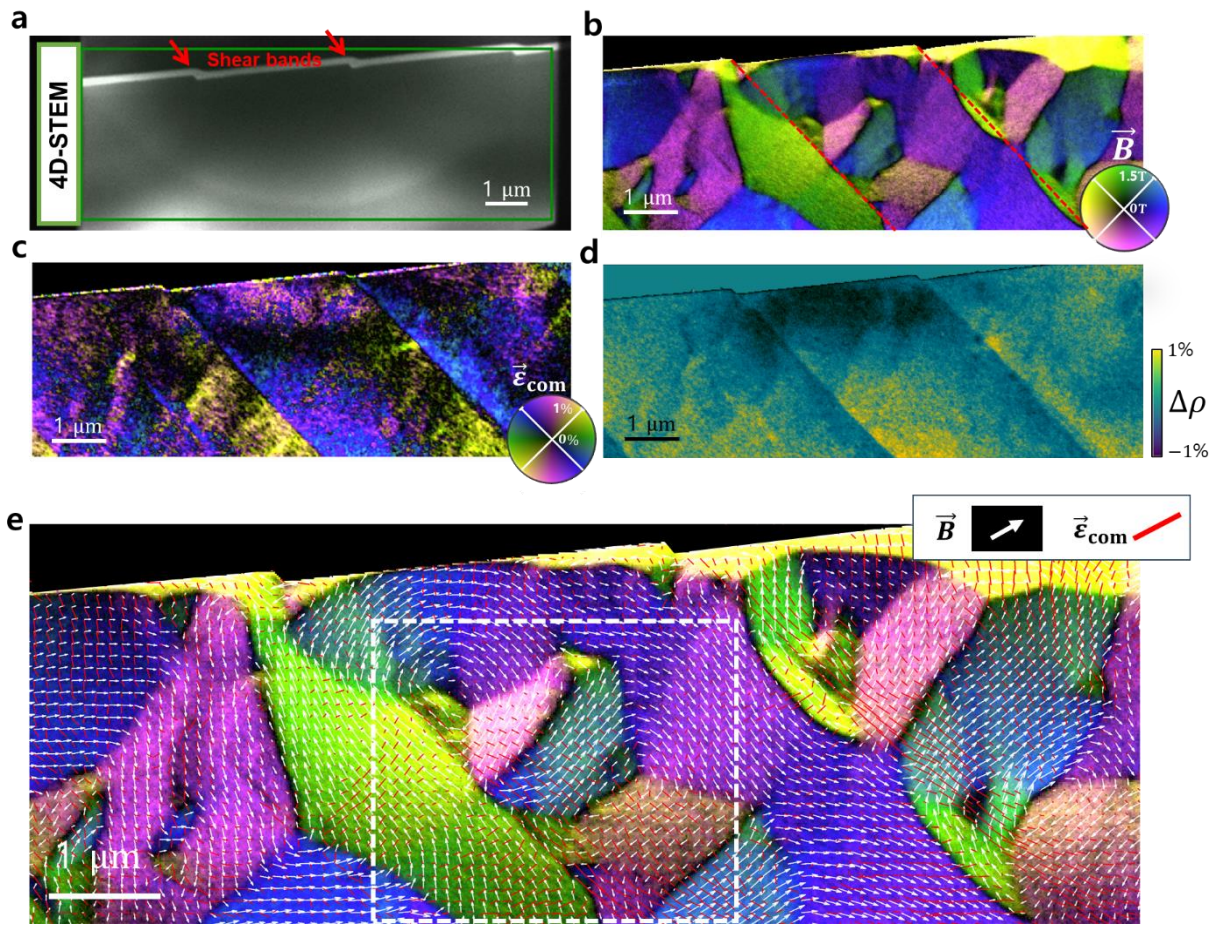

**Fig. 2:** Visualization of the magnetic field and atomic strain coupling using Ltz-4D-STEM observation of a plastically deformed amorphous metallic alloy. **a** STEM-ADF image of a TEM lamella, where the shear offsets are indicated by red arrows at the sample surface. A Ltz-4D-STEM map was acquired at the area indicated by the green rectangle. **b** Magnetic field ($\vec{B}$) image created using the principle of DPC imaging, but based on measuring the shift of the center of the 1st diffraction ring. The color corresponds to the orientation, and the brightness corresponds to the amplitude of the fields as indicated by the color wheel. The locations of two shear bands are indicated by red dashed lines. **c** Strain map visualizing the compressive strain ($\vec{\varepsilon}_{com}$). The strain orientation is presented by a two-fold symmetrical color wheel, the brightness corresponds to the strain amplitude. **d** relative density (Δρ). Yellow color represents high density and dark blue color low density. **e** Simultaneous visualization. The local $\vec{B}$ field is represented by white arrows and the local $\vec{\varepsilon}_{com}$ field is represented with red sticks. Both field representations are overlayed on the image shown in Figure 2**b**.

We used a deformed $Fe_{85.2}Si_{0.5}B_{9.5}P_4Cu_{0.8}$ (at.%) amorphous metallic alloy (metallic glass) with saturation magnetostriction of $l_s \sim +40\times10^{-6}$ [32, 33] as an example in this study, which receives significant attention owing to its soft ferromagnetism and high mechanical strength. [26, 34, 35] The amorphous metallic alloy was deformed by scratch testing under ambient conditions using a diamond tip. As a result, shear offsets occurred on the scratched surface. Figure 2a shows an ADF-STEM image of the TEM lamella, where the location of shear bands can be estimated based on the shear offset at the surface. We acquired a Ltz-4D-STEM map in the area (green rectangle) including the shear bands indicated by the red arrows, with a step size of 10 nm for balancing the field of view and data size.

Figure 2b-d shows typical results obtained from the Ltz-4D-STEM analysis: maps of the magnetic field ($\vec{B}$), compressive strain ($\vec{\varepsilon}_{com}$), and relative density (Δρ). The brightness in the $\vec{B}$ and the $\vec{\varepsilon}_{com}$ images represent the strength of the fields, and the colors represent the orientation of $\vec{B}$ and $\vec{\varepsilon}_{com}$. For Δρ, the yellow color represents high density, and the dark blue color low density. Complicated magnetic nanostructures are observed in the heavily deformed zones, i.e., the vicinity of the shear bands and the worn surface (Figure 2b). They are clearly different from the magnetic domain structure of the undeformed sample (Figure S4), which exhibits larger homogeneous domains ($> 3~\mu m \times 3~\mu m$).

Figure 2c visualizes that the strain concentrates in the vicinity of shear bands with an orientation difference of ~ 90° at each side of the shear bands. The asymmetrical strain fields with a sharp transition across shear bands in line with the previous strain observations in deformed amorphous metallic alloys. [23, 36, 37] The strain also induces the variation of the relative atomic density Δρ which reflects the net volume change due to the hydrostatic stress. Δρ suffers a sudden change from positive to negative across the shear plane (Figure 2d), namely the pop-in side (where the surface of the material was pressed down) suffers a mainly compressive strain, while on the opposite, the pop-out side (the material surface was pushed out) suffers tractive force that induces material dilatation. The strain field gradually fades out away from the shear bands in both compressed and tensile regions.

To rule out that density variations can introduce a phase shift of the exit electron wave (hence inducing a beam tilt), resulting in an artifact in the magnetic image, we conducted a conventional 4D-STEM measurement at the same sample position with the objective lens on (in a fully out-of-plane magnetized state due to the strong magnetic field generated from the objective lens). None of the contrast variations across the shear bands was visible in the DPC-like center of mass image as shown in Figure S5. This indicates that the local density gradients are too small to contribute any disturbance to the magnetic image and that the observed features in Figure 1b reflect the pure magnetic field of the sample.

Based on Figure 2 b-d, it can be seen that more and smaller inhomogeneous magnetic domains are present on the pop-in side of the shear band, in which strong compressive strain parallel to the shear band exists (strong tensile strain perpendicular to the shear band due to Poisson's effect). The magnetic domain structure in this region is confined by the surrounding orthogonal strain regions, leading to small and periodically arranged magnetic domains forming due to the strain-induced magnetic anisotropy that is perpendicularly aligned to the shear band orientation. In contrast, on the pop-out side of the shear bands, the tensile strain is mostly parallel to the shear bands and the magnetic domain structure is simpler with the magnetization now aligned parallel to the shear band direction. The closure domains extend over both regions to reduce the magnetic anisotropy contributions in the formed magnetic domain state. Similar magnetic domain structures have been discussed in samples with orthogonal magnetic anisotropies in soft magnetic amorphous thin films induced by ion irradiation [38] and locally induced stress variation [39].

Ltz-4D-STEM enables pixel-to-pixel correlative visualization and analysis of the magnetic and atomic structure, confirming the magnetic domain interpretation above. We plot both information together in Figure 2e. The white arrows represent the magnitude and orientation of the local $\vec{B}$ vectors. The red sticks representing $\vec{\varepsilon}_{com}$ visualize the strain field ($\vec{\varepsilon}_{ten}$ and

$\vec{\varepsilon}_{com}$ are perpendicular to each other. The arrows and sticks are overlayed on the magnetic field image shown in Figure 2b. Figure 2e shows that the orientation of $\vec{B}$ and $\vec{\varepsilon}_{com}$ are well correlated (namely, white arrows are close to being perpendicular to the red sticks) in a major part of the map, except at domain walls and vortexes. In close vicinity to the shear bands, the correlation between $\vec{B}$ and $\vec{\varepsilon}_{com}$ is uniformly close to ~ 90°, showing that the magnetization is dominated by the local strain.

For materials with isotropic magnetostriction properties, the magnetoelastic energy density can be written as $e_{ME} = -\frac{3}{2}\lambda_s \sum_{i=1}^{3} \sigma_i \gamma_i^2$, [40] where $\lambda_s$ is saturation magnetostriction, $\sigma_i$ is the deviatoric strain and its in-plane component can be quantified through the strain measurement (Figure S10a), $\gamma_i$ is the sine of the misorientation angle between $\vec{B}$ and $\vec{\varepsilon}_{ten}$. Taking advantage of the correlative imaging of the in-plane components of $\sigma_i$ and $\gamma_i$, it is expected that a map of $e_{ME}$ can be obtained if the strain and magnetic vectors were measured in 3D by involving the tomography strategy for vector field [41, 42] in our newly developed approach. As a simplified illustration, Figure S10b shows the map of the in-plane contribution of $e_{ME}$.

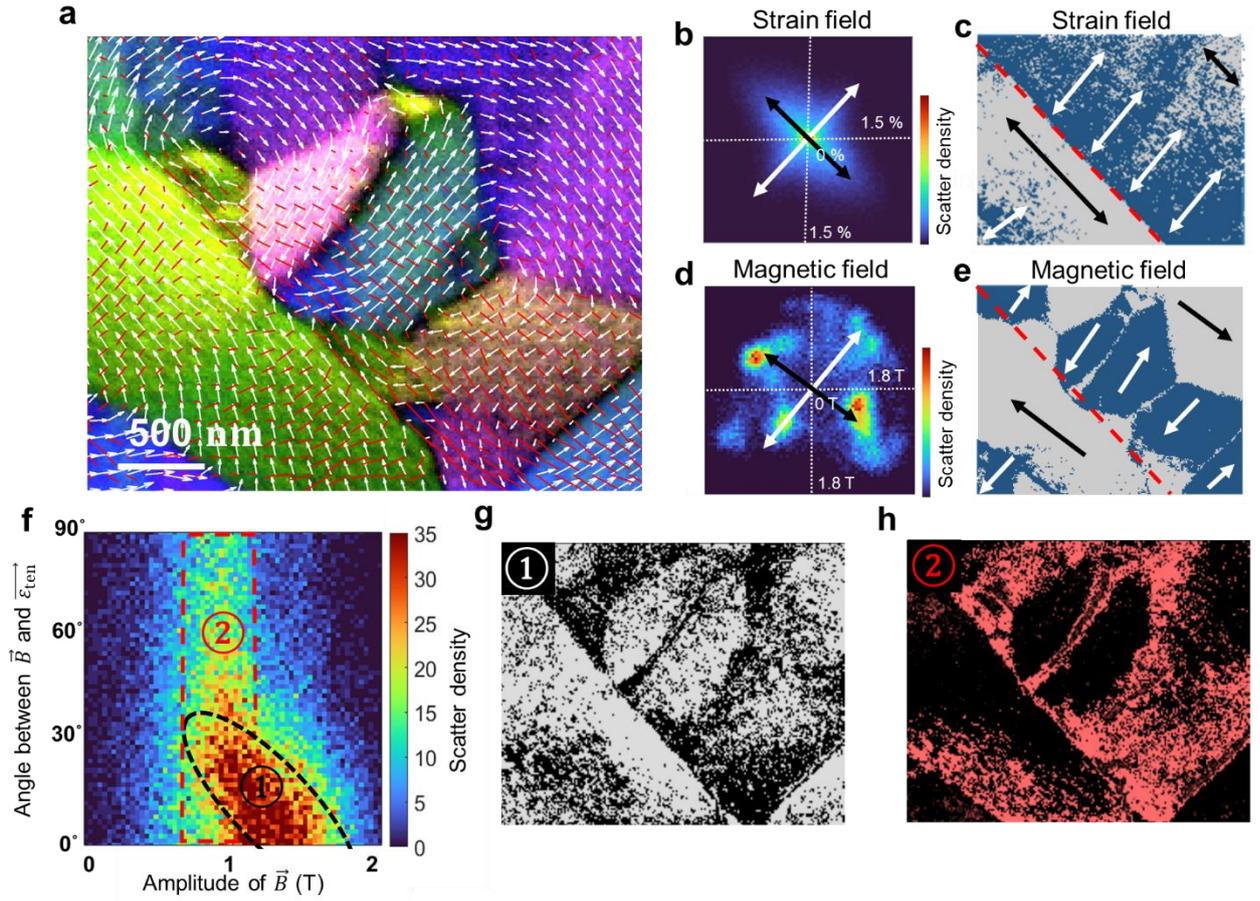

**Fig. 3:** Statistical analysis of the magnetic and strain coupling in the deformed amorphous metallic alloy using the Ltz-4D-STEM data. **a** Co-visualization of magnetic and strain field: a magnified image of the area indicated by the white dash rectangle in Figure 2a. **b** 2D histogram of $\vec{\varepsilon}_{ten}$. The x and y axes correspond to the horizontal and vertical components of the tensile strain measured at each pixel in the map and the color corresponds to the number of pixels. The color corresponds to the number of pixels. Black and white arrows highlight the principal orientations. **c** Binarized maps visualizing the spatial distribution of $\vec{\varepsilon}_{ten}$ following the two principal directions in **b**. Black and white arrows highlight the local major orientations. **d** 2D histogram of $\vec{B}$ following the same approach as in **b**. Black and white arrows highlight the principal orientations. **e** Binarized maps visualize the spatial distribution of $\vec{B}$ following the two principal directions in **d**. Black and white arrows highlight the local major orientations. **f-h** Correlative analysis of local strain and magnetization. **f** is a 2D distribution map of map pixels, in which the horizontal and vertical axes are $|\vec{B}|$ and the misorientation angle between $\vec{\varepsilon}_{ten}$ and $\vec{B}$. **g** and **h** are the spatial distribution of the map contained in the components ① and ② in **f**.

We statistically studied the magnetoelastic coupling in the highly strained regions around the shear band (Figure 3a). Figure 3b and d show 2D histograms of $\vec{\varepsilon}_{ten}$ and $\vec{B}$, where the axes correspond to the horizontal and vertical components of the tensile strain and magnetic moments measured at each pixel in the multi-information map (Figure 3a) and the color corresponds to the number of pixels counted for each map. Figure 3b reveals a two-fold symmetry of the strain field with the principal orientations along (black arrows) and orthogonal

(white arrows) to the shear band. Figure 3c shows the spatial distribution of the tensile strains binarized according to their orientation following the principal orientations shown in Figure 3b (blue is for perpendicular and light gray for parallel to the shear band). It reveals a 90° rotation of the strain field between the pop-in and pop-out sides across the shear plane. Figure 3d shows a clustered distribution of $\vec{B}$, where each cluster corresponds to a magnetic domain (Figure S6 shows the result from the whole map). The magnetic moments in the highly strained areas are preferentially orientated along the two principal strain directions, which are perpendicular to each other. Figure 3e shows the spatial distribution of the clusters (shown in Figure 3b) according to their orientations following the principal orientations. The magnetic domains are highly coherent with the strain field.

As shown above, Ltz-4D-STEM maps enable the analysis of multiple physical quantities. In the current study, these are the in-plane $\vec{B}$ field with two degrees of freedom and in-plane strain tensor with 3 degrees of freedom (symmetric 2×2 matrix, in the above results represented in the form of two perpendicular principal strains) which can also be interpreted in different aspects such as deviatoric strain and atomic density (i.e. the volumetric component). Owing to the pixel-to-pixel correlation between the maps of these different properties, there are many possibilities for correlatively analyzing these quantities as statistically meaningful. Figure 3f-h shows an example. Figure 3f is a 2D distribution plot (2D histogram) as a function of the amplitude of $\vec{B}$ and the misorientation angle between $\vec{\varepsilon}_{\text{ten}}$ and $\vec{B}$. The major population of the pixels concentrates at the small misorientation angle (< 30°), forming cluster ① in Figure 3f. This reveals a preference that the in-plane component of $\vec{B}$ increases linearly with decreasing misorientation angle between $\vec{\varepsilon}_{\text{ten}}$ and $\vec{B}$. This suggests that higher in-plane tensile strain tilts the magnetization in-plane, resulting in a stronger strength of the projected magnetic field to reduce the magnetoelastic energy. Analyzing the population in Region ② in Figure 3f, $\vec{B}$ show strong independence to the strain orientation. Figure 3g and h visualize the spatial distribution corresponding to Regions ① and ②. Region ① coincides with the primary domain regions which are well-defined domains with a lower angular mismatch between the magnetic and strain fields so that minimizes the magnetoelastic energy. Region ② contains the closure domains with more complex magnetic structures e.g. domain walls and vortexes. The amplitude of the in-plane magnetic field in Region ② is weaker than the primary domains in Region ① (as visualized in Figure S5a).

More examples of correlative analysis across the multi-physical quantities are shown in Figure S7 and S8, as well as a direct correlative analysis crossing a domain wall as shown in Figure S9.

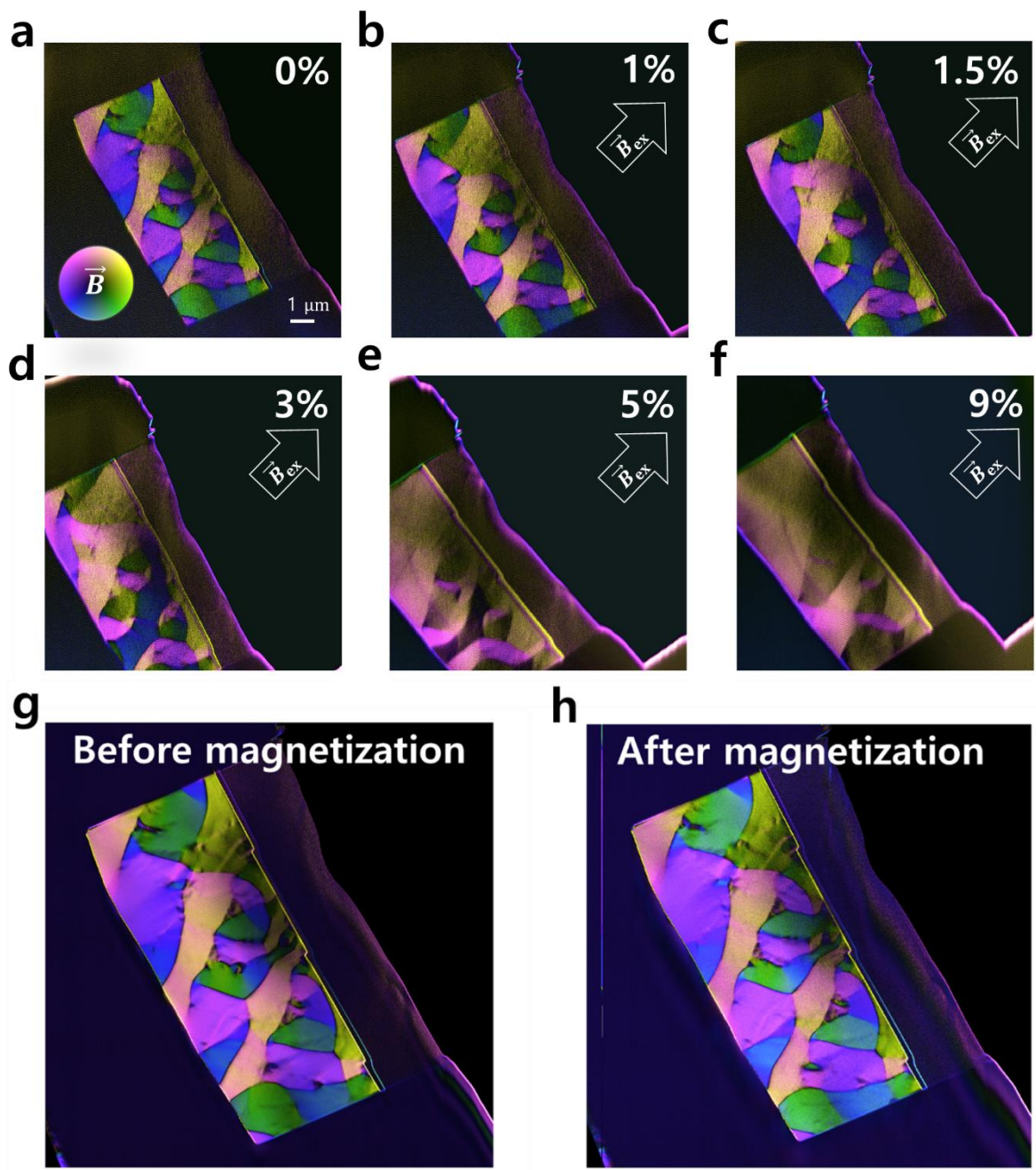

**Fig. 4:** *In-situ* magnetization test of a plastically deformed amorphous metallic alloy by activating the objective lens. The magnetic field is imaged using a 4-segment detector in a conventional DPC setup. The sample is tilted to 10 degrees to apply an in-plane magnetic field to the TEM lamella. The objective lens was exited from **a** 0 % to f 9 %, corresponding to the strength of the magnetic field in the sample plane 0 to 30 mT. The inset arrows indicate the in-plane direction of the external magnetic field $\vec{B}_{ex}$. **g** and **h** are high-quality maps before and after *In-situ* magnetization test recorded with longer exposure time.

Figure 4 shows an *in-situ* magnetization test for the same TEM sample (supplementary video 1). The lamella was tilted to 10 degrees to apply a true in-plane magnetic field to the sample. The objective lens is gradually excited from 0 % to 9%, corresponding strength of the magnetic field in the sample plane is from 0 to 30 mT (estimated by the fact that the 100% objective lens produces a 2 T magnetic field). Although the major magnetization in the sample is oriented along the external magnetic field ($\vec{B}_{ex}$), the complex domain pattern formed due to the local strain field is strongly resistant to change compared to the situation of the undeformed sample (supplementary video 2). We note that the magnetic domain structure almost recovered back to the original magnetization state after the objective lens current was set back to zero at the end of the test (Figure 4g and h), except for only slight variations in some domain walls and vortexes. This indicates very low magnetic hysteresis in the material. The magnetic domain patterns are determined by the underlying strain pattern, namely the magnetoelastic anisotropy contributions in the soft magnetic material.

**Discussion**

The Ltz-4D-STEM method fills the gap of simultaneously mapping both the atomic and magnetic domain structure in a material. The obtained dataset contains spatial information of multiple physical properties fully correlated at the pixel level. Using this, the magnetoelastic coupling was directly visualized. The field of view can easily be extended up to tens of micrometers and is only limited by the sample size and the volume of the hardware data storage. The highest possible scanning resolution of the map is defined by the probe size which is limited by the spherical aberration of the probe forming lens. For the probe-aberration corrected system in Ltz-STEM condition, it was demonstrated to be less than 1 nm. Due to the reduced camera length, the sensitivity of magnetic field measurements would be lower compared to when only the transmitted center beam is used. However, as the magnetic field does not only deflect the center beam but also the whole diffraction pattern (Figure S2c), we used the center of the 1st diffraction ring to generate a measurement analogous to the DPC signal, which shows very high accuracy and sensitivity to measure the magnetic field induced beam deflection (~ 0.1 µrad) and strain (0.08 %).

The deviatoric strain gives rise to the development of structural anisotropy dominating the magnetoelastic coupling. [40, 43] As the Fe-based amorphous metallic alloy possesses positive saturation magnetostriction ($\lambda_s$~40 ppm), the magnetic moments tend to be aligned parallel to the tensile strain direction and perpendicular to the compressive strain direction. [44, 45] The local strain tends to reorient the local magnetic moments. [11-13, 43, 44, 46] to minimize the magnetoelastic energy, namely the magnetic moments preferably align to the direction of strain-induced magnetic anisotropy. Therefore, the magnetic microstructure is strongly determined by the residual elastic strain field. For magnetically soft ferromagnetic materials, the physical size of the domains is usually large to minimize the density of domain walls. [7] However, a high (magnetostatic) energy configuration was observed around the shear bands: the pop-in side of the shear band contains a number of small domains oriented towards the shear band, whereas the other side of the shear band contains mostly a single domain-oriented along the shear band. In such highly strained areas (e.g. in Figure 3a), the magnetization and strain are well correlated within the primary domains (Figure 3g), but in the regions of the closure domains and vortexes (Figure 3h) the magnetoelastic coupling seems to be broken. This is due to that the magnetic configuration is governed by both the magnetic anisotropy (magnetoelastic) and the dipole-dipole interaction (magnetostatics). Due to the sharp 90° transition of the strain states crossing the shear bands, the magnetic moments close to shear bands suffer strong competition between the magnetoelastic and magnetostatic energy, which is responsible for the formation of complex magnetic structures.

Interestingly, the amplitude of the in-plane magnetic components exhibits significant variations in different domains (Figure S5a). The exact reasoning is not clear from our images, but this observation suggests the additional out-of-plane magnetization contributions to reduce the magnetization energy in the small bar-shaped sample. In particular, the observed complexity in the closure domain regions infers a possible supplementary or more complicated magnetic structure varying in the sample thickness direction due to additional flux closure between the submicron domains. This may lead to the observed reduction of the net in-plane magnetic flux density in some domains. We also cannot rule out the possibility of out-of-plane strains, which may also be responsible for the existence of out-of-plane magnetization components.

The strain-induced anisotropy pattern leads to a high reproducibility of the magnetization patterns at the nanoscale as observed in the *in-situ* magnetization test. This may lead to potential applications for nanomagnetic devices. [45, 47, 48] The method can be applied to both amorphous and crystalline magnetic materials, Figure 1b (bottom) shows an example of the nanobeam electron diffraction pattern of a SmCo magnet recorded in the field-free Lorenz optical mode, where the crystal symmetry, orientation, and phase analyses can be performed using the 4D dataset. Of course, the magnetic field imaging with this method for crystalline materials will be expected to suffer from dynamic electron scattering issues as in conventional field-free DPC when the sample is not thin enough. Applying beam precession has been shown to significantly reduce these artifacts. [49, 50]

As demonstrated, the advanced field-free optics and 4D-STEM strategy can simultaneously image the magnetic field and atomic structures with a few nanometers probe size, sub-angstrom atomic structure information, and across an extensive field of view. This approach breaks the information limits of TEM in studying nanostructured magnetic materials and enables direct visual correlation of atomic structure and local magnetic properties.

**Experiments**

$Fe_{85.2}Si_{0.5}B_{9.5}P_4Cu_{0.8}$ (at.%) master alloy ingots were prepared from the melt by rapid solidification on rotating Cu wheels at Vacuumschmelze GmbH & Co. KG resulting in a ribbon width of about 25 mm and a thickness of about 20 µm. Scratch tests were performed with a scratch length $l_s = 1$ mm with normal loads $F_n = 10$ N for the Fe-based metallic glass ribbon with a sliding velocity $v_s = 0.1$ mm/s at ambient conditions using a diamond tip with a radius of 210 µm. For the STEM investigation, TEM lamella was prepared by FIB (FEI Strata 400S) from the ribbon at the left side of the scratch and 0.3 mm away from the scratch end. Thinning was performed to a sample thickness of about 200 nm for electron transparency at an acceleration voltage of 30 kV with gradually decreasing beam currents from 8 nA to 2 pA to reduce the ion beam damage.

4D-STEM measurements were conducted using a Themis Z double-corrected TEM (ThermoFisher Scientific) operated at 300 kV in microprobe STEM mode at field-free condition with the Lorentz lens activated with spot size 6 and a semi-convergence angle of

0.26 mrad giving rise to a probe size of ~5 nm. The field-free condition was checked by observing the domain structure of a TEM lamella lift-out from the undeformed area of the Fe-based amorphous metallic alloy. For 4D-STEM data acquisition, we used a OneView camera (Gatan Inc.). The magnification of the diffraction pattern (i.e. camera length) was set to capture the first diffraction ring with a large diameter spanning > 80% range of the camera to enhance the sensitivity and accuracy for measuring the distortion. 4D-STEM maps were acquired by scanning the electron probe over a 2D sample plane with a step size of 15.8 nm and a frame size of 620×225 and camera binning to 256×256 pixel$^2$ for balancing the storage space in our camera computer with an exposure time of 3.3 ms per frame (frame rate of ~300 f/s). The shift of the central position of the diffraction pattern during scanning was eliminated by careful alignment of the scan pivot point and Descan function. This is benefited from the relatively long distance between the sample and condenser/Lorentz lenses. A remaining diffraction shift can be corrected by post-background processing by capturing a reference map without the sample.

The diffraction pattern of a typical amorphous material shows a diffuse ring pattern (Figure 1a). The magnetic fields inside the sample deflect the electron beam due to Lorentz force. The position of the direct beam is a measure of the direction and strength of the local magnetic field, which contains the 1$^{st}$ order gradient of the phase of the electron wave exiting from the specimen, creating so-called differential phase contrast (DPC).

The local stress in the metallic glass induces a structural anisotropy, which results in an elliptic distortion of the diffraction ring leading to a deviation from the ideal circle as illustrated in Figure 1b (The diffraction pattern was further elongated artificially for easy presentation). Therefore, the strain can be mapped by determining the ellipticity of the diffraction ring in each local diffraction pattern of the Ltz-4D-STEM dataset. The diffraction patterns are fitted with an algebraic method as illustrated in Ref. [23]. From the fitted ellipse, The relative density was determined as $\Delta\rho = \frac{q_{max}q_{min} - q_0^2}{q_{max}q_{min}}$ and the principal strains were determined as $\vec{\varepsilon_1} = \frac{q_0 - q_{max}}{q_{max}} \begin{pmatrix} \cos(\theta) \\ \sin(\theta) \end{pmatrix}$ and $\vec{\varepsilon_2} = \frac{q_0 - q_{min}}{q_{min}} \begin{pmatrix} \cos(\theta + 90°) \\ \sin(\theta + 90°) \end{pmatrix}$, where $q_0$ is the radius of the 1$^{st}$ ring for the unstrained case (averaged from an area far away from the deformed region) and $q_{max}$ and $q_{min}$ are the lengths of the maximum and minimum elliptical axes of the 1$^{st}$ ring. $\theta$ is the corresponding azimuthal angle between $q_{max}$ and the x-axis. The deviatoric strain was calculated from both principal strains as $\varepsilon_{dev} = \frac{|\vec{\varepsilon_1}| - |\vec{\varepsilon_2}|}{2}$.

**Acknowledgments**

The authors express their gratitude to the Karlsruhe Nano Micro Facility (KNMFi) for their assistance and providing access to FIB and TEM facilities. X. M. acknowledges financial support from the Deutsche Forschungsgemeinschaft (DFG) funding (MU 4276/1-1). Additionally, the authors appreciate the support received from the Joint Lab Model Driven Materials Characterization (MDMC) and acknowledge the backing from the Helmholtz Imaging Project (HIP) BRLEMM.

**Author contributions**

S.J.K and X.M developed the methods and performed the TEM experiments. K.D and C.M provided the Fe-based sample. S.J.K and A.C performed the deformation experiments. S.J.K, X.M, D.W, R. D.-B., J.M., and C.K analyzed and discussed the data. X.M and C.K supervised the project. S.J.K, X.M, and C.K wrote the manuscript. All authors contributed to the revision of the manuscript.

**Competing interests**

The authors declare that they have no competing financial interests.

**Data availability**

Data is available on open access via KITOpen and also on request from the authors.

Correspondence and requests for materials should be addressed to X.M. and C.K (e-mail: xiaoke.mu@kit.edu and christian.kuebel@kit.edu )

**Supplementary**

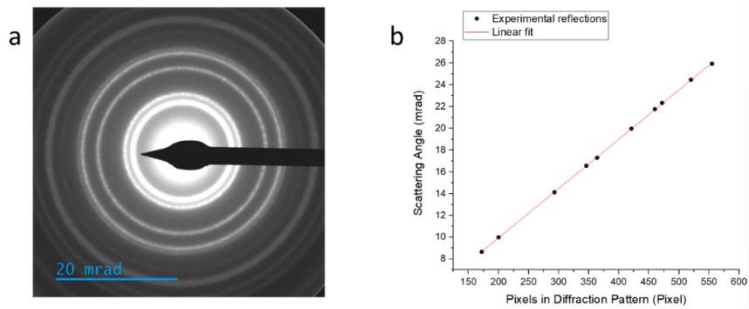

**Figure S1:** (a) Diffraction pattern of a standard cross-grating sample in Ltz-4D-STEM mode with reduced C3 lens for creating a large beam size and enough sampling of Au nanoparticles. (b) The diameter of the Au diffraction rings measured on the camera in units of number of pixels versus their theoretical scattering angle for 300 keV electron. The black dots are experimental data and the red line is a linear fit to the experimental data.

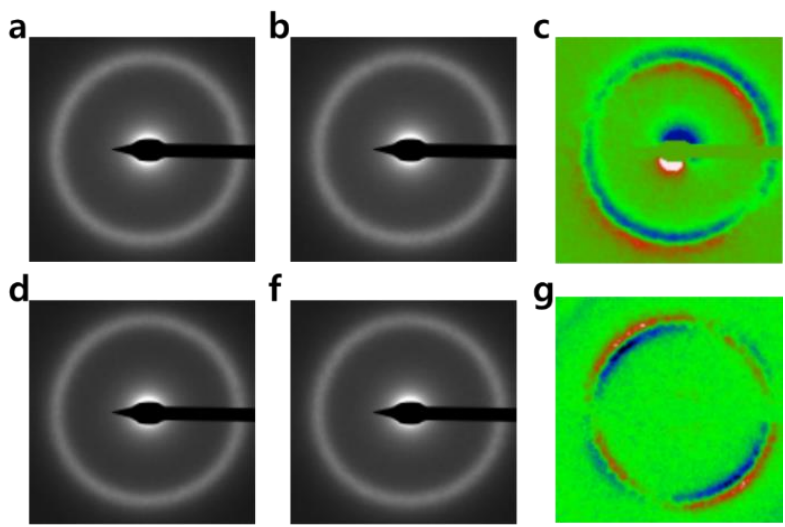

**Figure S2:** Diffraction patterns acquired by Ltz-4D-STEM from different sample locations. (a) From a domain pointing left-up direction and (b) right-down direction. (c) Subtraction of both (a) and (b) diffraction patterns. (d) Unstrained case, and (f) strained case. (g) Subtraction of both (d) and (f) diffraction patterns

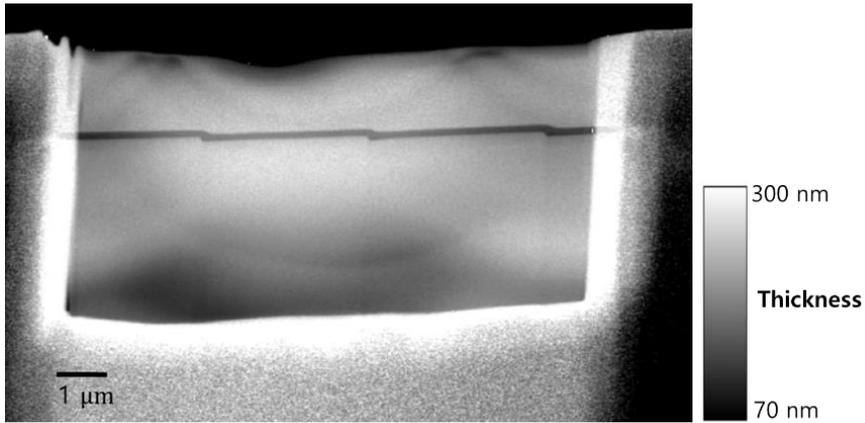

**Figure S3:** Thickness map of the FIB-prepared lamella from the deformed Fe$_{85.2}$Si$_{0.5}$B$_{9.5}$P$_4$Cu$_{0.8}$ amorphous metallic alloy obtained using energy-filtered transmission electron microscopy (EFTEM). The mean free path (MFP) of an electron for the material was estimated to be ~75 nm for the thickness estimation.

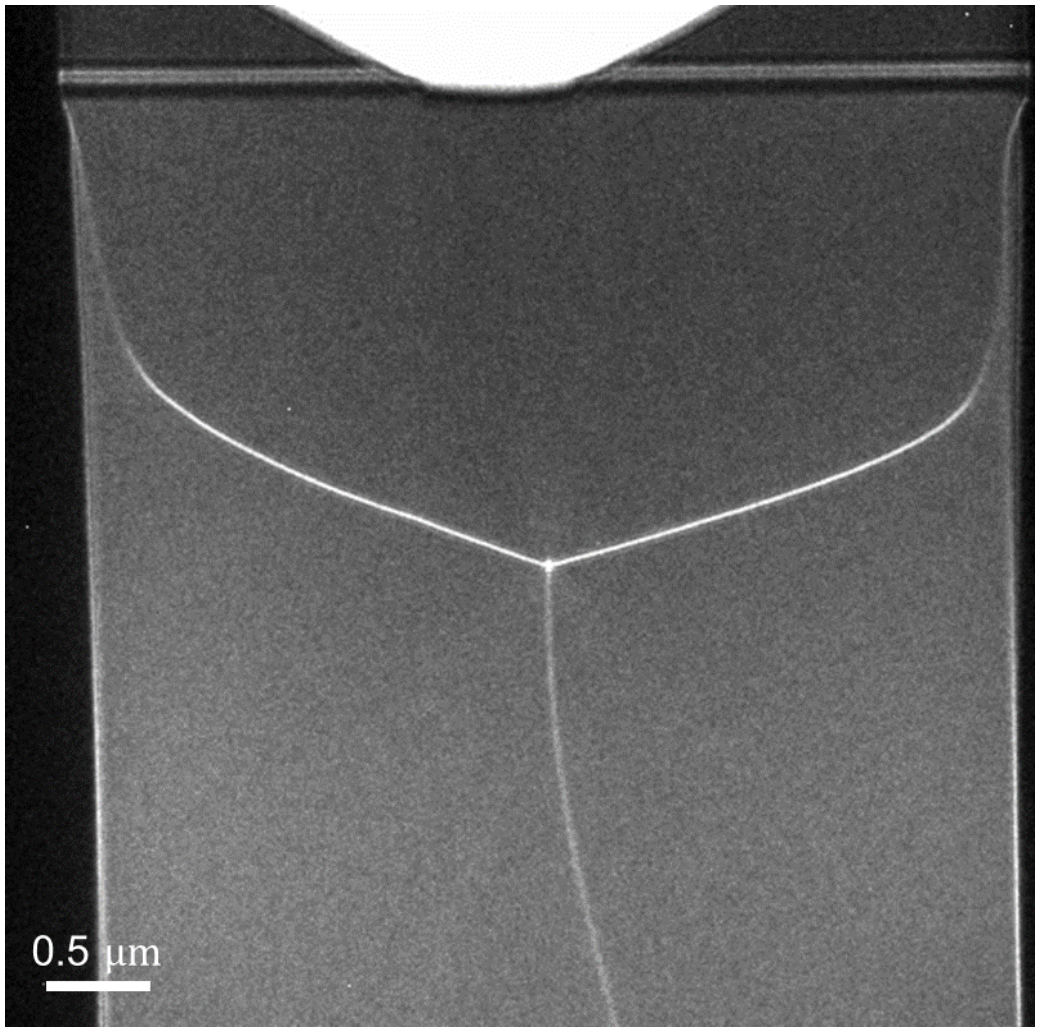

**Figure S4:** Ltz-TEM images of two TEM lamellae of the undeformed Fe$_{85.2}$Si$_{0.5}$B$_{9.5}$P$_4$Cu$_{0.8}$ amorphous metallic alloy showing the domain structure of the soft ferromagnetic material.

Notably, the undeformed sample may also possess some degree of residual stress anisotropy from fast melt-spin quenching, leading to the observed variation of the magnetic domain structure from a regular Landau-like domain structures, reflected in the curvature of the domain walls and their variation from a regular closure domain angle.

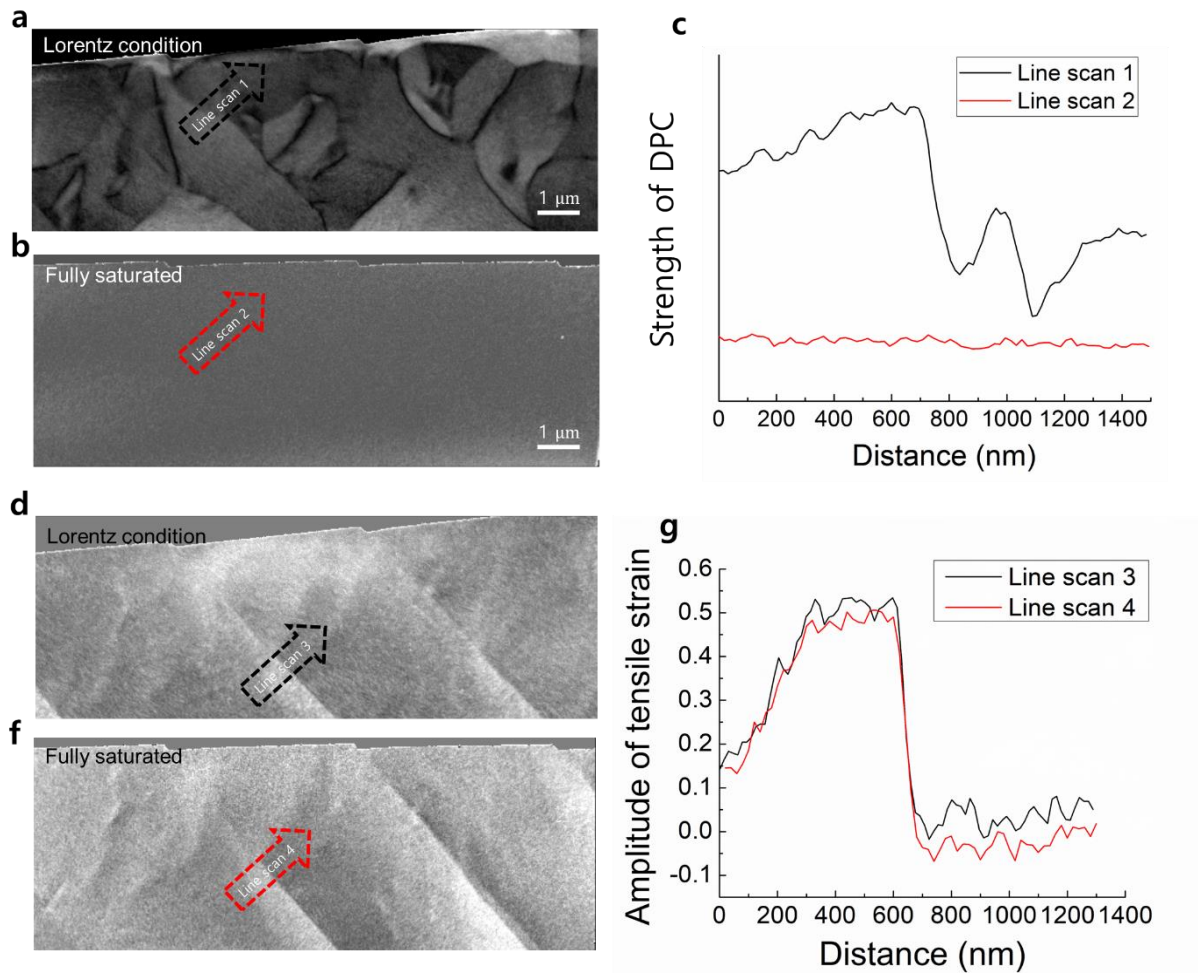

**Figure S5:** Map of the amplitude of the in-plane magnetization of the deformed amorphous $Fe_{85.2}Si_{0.5}B_{9.5}P_4Cu_{0.8}$ metallic alloy at **a** unmagnetized state (field-free condition), and **b** fully vertically magnetized state (conventional microprobe STEM condition with objective excited about 2 T). **c** Line profiles taken across the shear band. Line scan 1: From the image of **a** along the black dash arrow. Line scan 2: From **b** along the red dash arrow. Map of the amplitude of the tensile strain at **d** unmagnetized state (field-free condition), and **f** fully vertically magnetized state. **g** Line profiles taken across the shear band. Line scan 3: From the image of **d** along the black dash arrow. Line scan 4: From **f** along the red dash arrow.

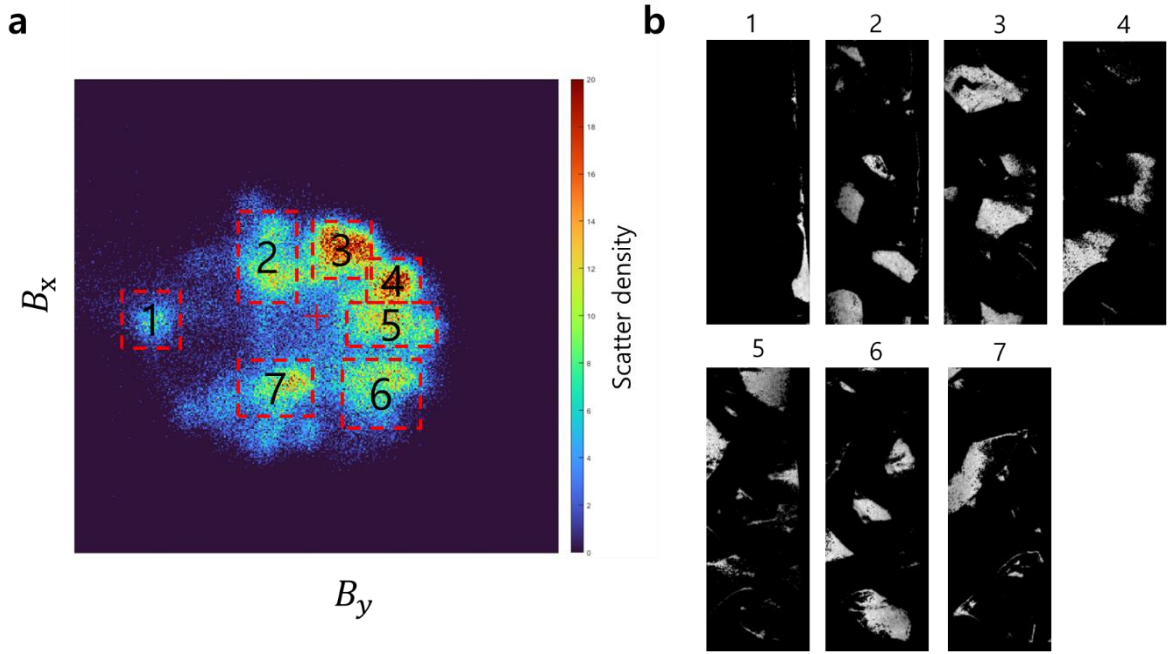

**Figure S6: a** 2D Histogram plot of vector components of the magnetic field ($B_x$ vs. $B_y$). over the whole mapping area. The dash rectangles with numbers 1 to 7 are used for mapping. **b** Corresponding maps from the selections in **a** are shown.

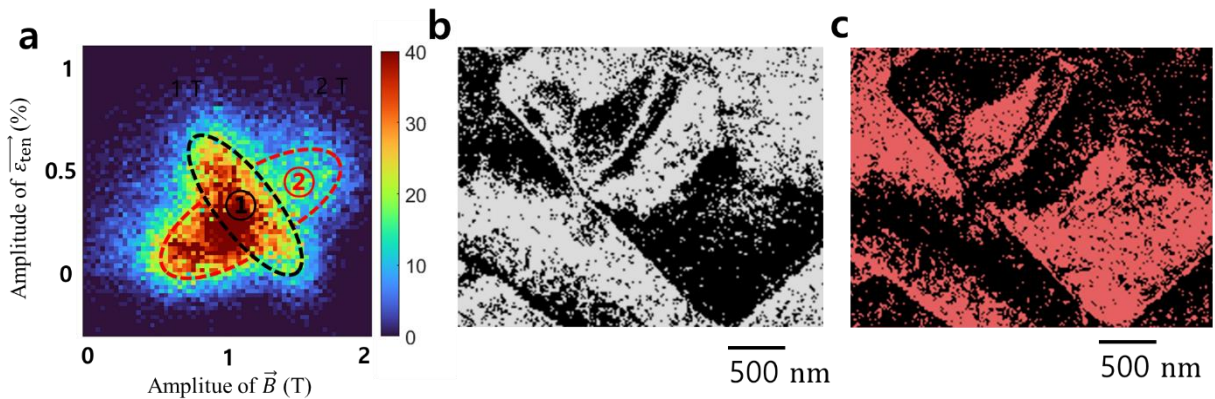

**Figure S7:** Correlative analysis of local strain and magnetization for a plastically deformed amorphous metallic alloy. **a** 2D distribution map of pixels according to $|\vec{B}|$ and $|\vec{\varepsilon}_{ten}|$. It shows an interesting star-shaped feature with proportional (region ②) and inverse proportional (region ①) correlations. **b** and **c** Spatial distribution of map pixels contained in the components ① and ② in **a.**

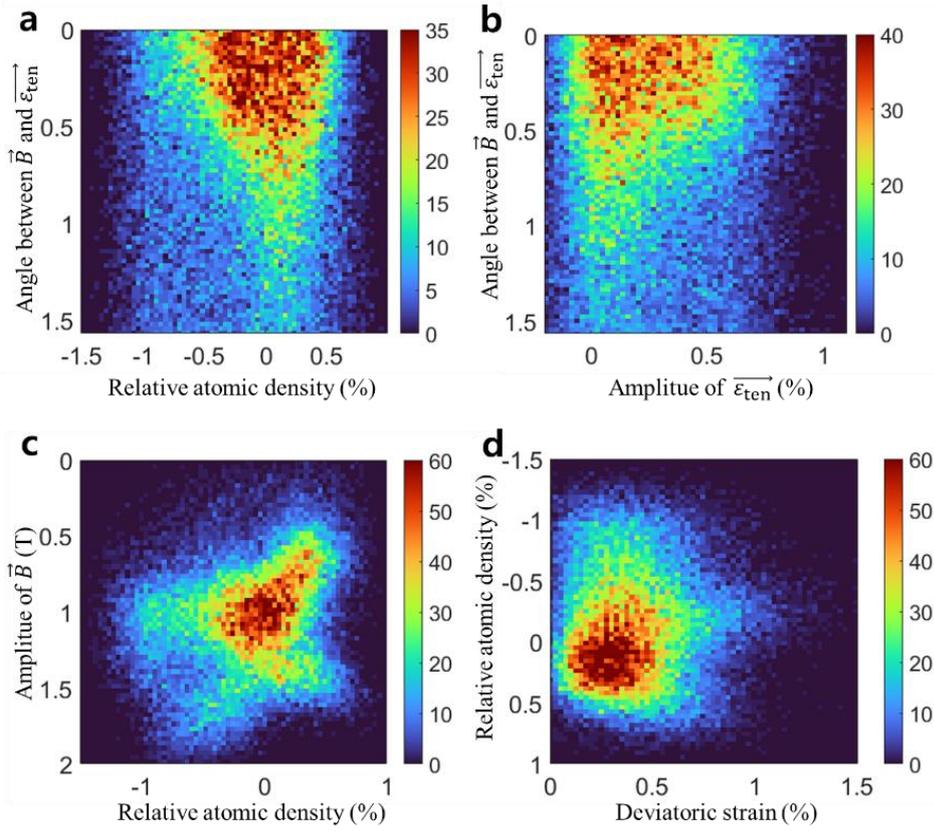

**Figure S8:** Exemplary density plots (2D histogram analysis) of (a) misorientation angle of $\vec{\varepsilon}_{ten}$ and $\vec{B}$ and relative atomic density, (b) misorientation angle of $\vec{\varepsilon}_{ten}$ and $\vec{B}$ and amplitude of $\vec{\varepsilon}_{ten}$, (c) amplitude of $\vec{B}$ and relative atomic density, and (d) relative atomic density and deviatoric strain.

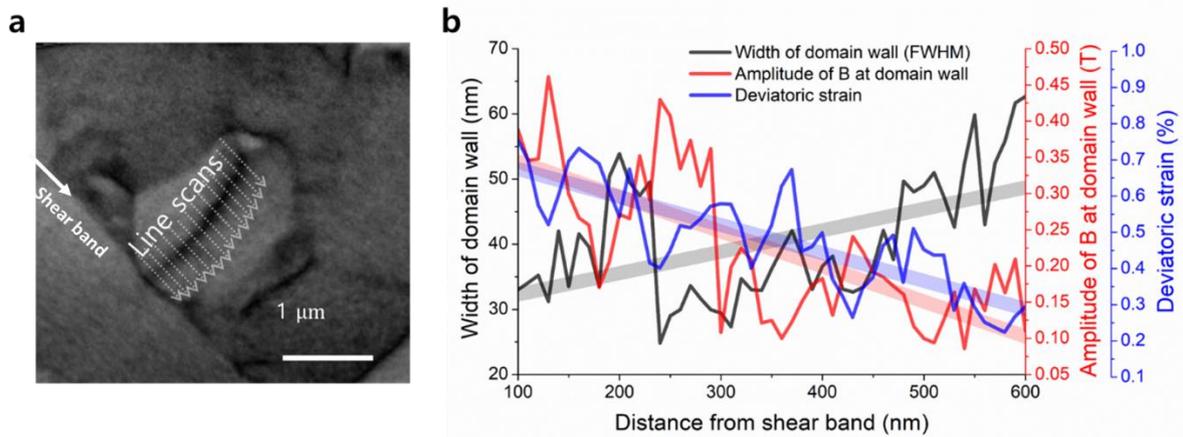

**Figure S9:** Correlative analysis across a domain wall. (a) An area of interest in $|\vec{B}|$ map. (b) Line profiles of the quantities of $|\vec{B}|$, $|\vec{\varepsilon}|$ and deviatoric strain, taken across the domain wall.

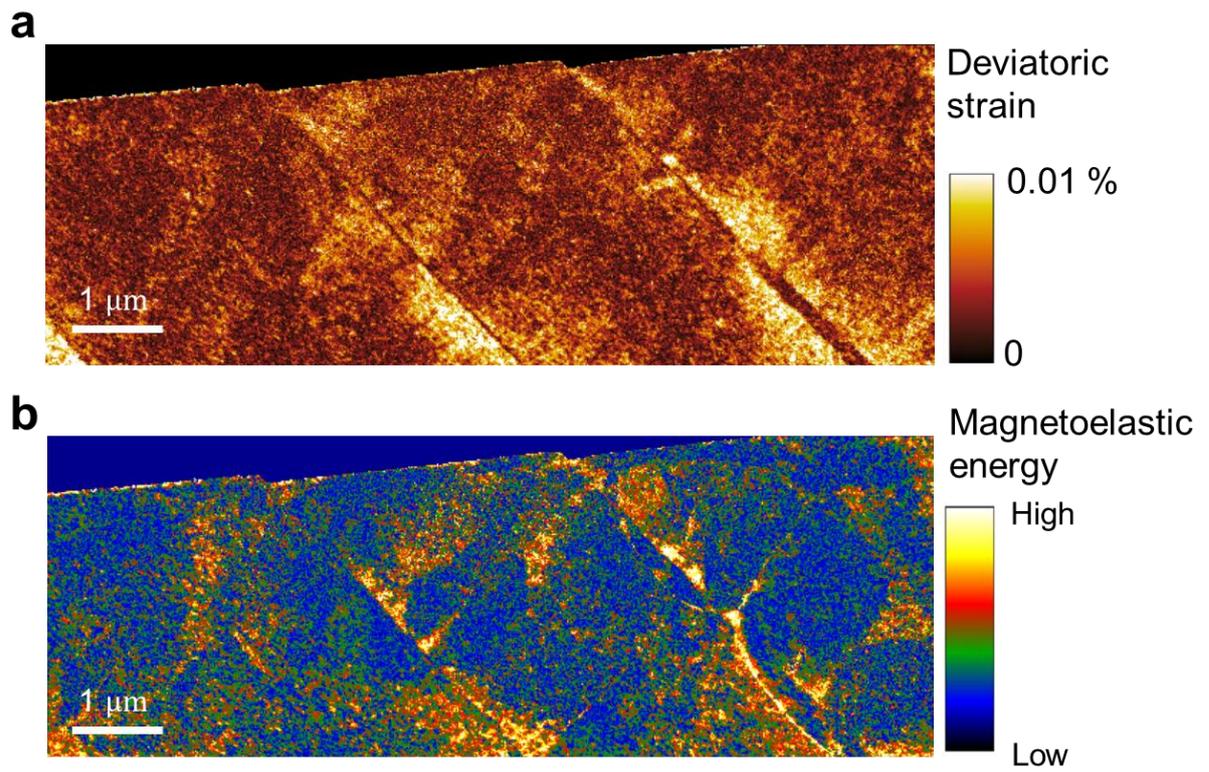

**Figure S10: a** Map of the deviatoric strain of the plastically deformed amorphous metallic alloy. **b** Map of magnetoelastic energy.